\documentstyle[11pt,paspconf]{article}

\begin{document}

\title{Small and Very Small Interstellar Grains}
\author{Adolf N. Witt}
\affil{Ritter Astrophysical Research Center, The University of Toledo, 
    Toledo, OH 43606, USA}

\begin{abstract}

	This review summarizes the observational characteristics of those interstellar grains, which are prevented from entering the solar system by interactions with the heliopause. Such grains are typically less than 100 nm in radius and they reveal their presence by interstellar absorption at ultraviolet wavelengths and by non-equilibrium emissions in the red and near-infrared portions of the spectrum.

\end{abstract}


\section{Introduction}

		The detection and mass-characterization of small solid particles, 
originating in interstellar space and now entering and traversing the
solar system, has been a major accomplishment achieved with the current Ulysses
and Galileo space probes (Gruen et al. 1994; Frisch et al. 1999). However, only
the largest interstellar grains are actually detected with these experiments, because the 
interaction between charged grains and magnetic fields at the heliopause serves as an 
effective discriminator for entry into the heliosphere for grains smaller than about 100 nm 
radius, which is expected to lead to a significant overabundance of the excluded grains at 
the heliopause. This review presents the astronomical evidence for the existence of a large 
population of small and very small grains in the diffuse interstellar medium of the Galaxy, 
a population subject to exclusion at the heliopause (Frisch et al. 1999).

\section{Definition of Size Domains of Interstellar Grains}
		Various interactions in interstellar space involving grains and grains, 
grains and gas, and grains and photons and other energetic particles,
e.g. coagulation, shattering, accretion, sputtering, and photodestruction,
in addition to the influx of newly formed grains from stellar mass outflows, assure the 
existence of a wide, very likely continuous, distribution of sizes of solid particles, 
generally referred to as interstellar grains. Recent determinations of this size distribution can be found in the work of Kim et al. (1994) and of Zubko et al. (1999). For operational reasons it makes sense to 
divide this distribution into three separate domains:
\begin{enumerate}
\item "typical" interstellar grains
\item "small" interstellar grains (SG)
\item "very small" interstellar grains (VSG)
\end{enumerate}
The approximate boundaries in radius (assuming spherical grains) and mass separating these 
domains as well as the relative abundance of the respective grains are delineated in Table 1.

\begin{table}
\begin{center}
\caption{Size Domains of Interstellar Grains. \label{tbl-1}}
\begin{tabular}{cclc}
\tableline\tableline
Domain	& Radius (nm)	& Mass (g)      & Relative Abundance \\
\tableline
typical & 100 - 1000 & $10^{-14}$ - $10^{-11}$ & 1	\\
small (SG) & 5 - 100 & $10^{-19}$ - $10^{-14}$ & $\sim10^{3}$ \\
very small (VSG) & 0.3 - 5 & $10^{-21}$ - $10^{-19}$ & $\sim10^{6}$ \\
\tableline
\end{tabular}
\end{center}
\end{table}

Since "typical" interstellar grains will not be dealt with in detail in subsequent parts of 
this review, a brief summary of their observational characteristics is given now. "Typical"
grains penetrate the heliopause and 
are recognizable by dust detectors on spacecraft by virtue of their heliocentric velocity 
vector; they are the principal contributors to extinction of star light in interstellar 
space at visible wavelengths, with an approximate $\lambda^{-1}$ wavelength dependence; 
they are 
responsible for scattering of starlight in the interstellar medium and in reflection 
nebulae; their efficient alignment by interstellar magnetic fields, giving rise to 
interstellar polarization of star light, is testimony to their non-sperical shapes; their 
average temperature of 17$\pm$2 K derived from absorption of the typical interstellar 
radiation
field, characterizes the far-infrared emission component in the electromagnetic spectrum of 
the Milky Way galaxy as well as other dust galaxies; and finally, despite their relatively 
small number density, the typical interstellar grains make up the bulk of the mass found in 
solid particles in the diffuse interstellar medium. A good summary of our current knowledge 
of typical interstellar grains can be found in the monograph by Whittet (1992).

\section{Observational Characteristics of Small and Very Small Interstellar Grains}

\subsection{Extinction}
		The extinction characteristics of SGs and VSGs can be derived from Mie 
theory (e.g. Bohren \& Huffman 1983) in the limit where the size parameter
\begin{eqnarray}
                            x  =  \frac{2 \pi a}{\lambda}~~<< 1         
\end{eqnarray}
where a is the particle radius. One finds, in this case, that the absorption efficiency 
$Q_{abs}$ is directly proportional to x and the scattering efficiency, $Q_{sca}$, is negligibly 
small. As a result, the extinction is determined entirely by the absorption cross section
\begin{eqnarray}
C_{abs} = \pi a^2 Q_{abs}
\end{eqnarray}	
and, hence, is proportional to the volume (or mass) of the SGs and VSGs and inversely 
proportional to the wavelength. Consequently, at the shorter  ultraviolet wavelengths, where
the extinction by the larger "typical" grains has reached saturation, the presence of SGs 
and VSGs reveals itself through a continued wavelength-dependent extinction increasing as 
1/$\lambda$ with decreasing wavelength. This is indeed a general characteristic of the 
observed interstellar
extinction, leading SGs and VSGs to absorb about 35\% to 40\% of the energy in the diffuse 
interstellar radiation field in the Galaxy.
Unfortunately, in the limit set by Eqn. (1), the UV extinction curve is not
particularly sensitive to the sizes of SGs or VSGs, but the extinction in the UV is a good 
measure of the total mass fraction of SGs and VSGs. As a consequence, competing models for 
interstellar grains fit the extinction observations equally well, assuming either a 
bi-modal size distribution (e.g. Greenberg et al. 1973)
or a continuous power-law distribution (e.g. Mathis et al. 1977).

		The exploration of UV extinction along numerous lines of sight
with the International Ultraviolet Explorer (IUE) has revealed that the UV extinction in 
about 50\% of all cases deviates significantly from the Galactic average, indicating that 
the relative mass fraction contained in SGs and VSGs relative to "typical" grains varies 
considerably from one line of sight to another (Fitzpatrick 1999). Possible processes which 
might account for these variations are accretion of SGs onto "typical" grains, which would 
lead to a flat extinction in the UV, selective destruction of SGs, leading to the same 
result, or the shattering of large grains into SGs and VSGs, producing a steep rise in UV 
extinction. A remarkable study by Boulanger et al. (1994) of 8 stars in the Chamaeleon 
cloud complex has shown that extreme cases of extinction curves can occur simultaneously 
within a relatively small volume of space, a testimony to the efficiency of the processes 
altering the size distribution of grains in space.

\subsection{Scattering in the UV}

		Interstellar grains are highly efficient scatterers in the 
visible as well as the UV. The scattering characteristics of grains can be described in 
terms of the wavelength-dependent values of the albedo and the phase function asymmetry g, 
defined as the average cosine of the scattering angle. Grains comparable in size to the 
wavelength scatter strongly in the forward direction, resulting in g-values in excess of 
0.5. Studies of the phase function asymmetry in the UV have shown that g-values are 
increasing with decreasing wavelength (Witt et al. 1992; Calzetti et al. 1995), consistent 
with a case where only large particles are scattering. As pure absorbers, SGs and VSGs 
affect the derived scattering properties only in that they reduce the effective dust albedo 
in wavelength regions where enhanced absorption by these
grains occurs. Calzetti et al. (1995) demonstrated conclusively that the strong
2175 \AA\ feature in the UV is due to pure absorption, most likely due to VSGs,
and the steep rise in the far-UV extinction shortward of 1300 \AA\ is equally associated with a
decreasing albedo and hence is evidence for strong VSG absorption (Witt et al. 1993). Also,
the far-UV albedo derived from observations of scattering in the diffuse interstellar dust 
at high galactic latitudes, which is characterized by generally steeper far-UV extinction 
with decreasing wavelength, was found to be 50\% lower than the UV-albedo of dust in 
reflection nebulae, where dust particles on average are larger and the far-UV extinction is 
less steep (Witt et al.1997). The fact that scattering properties do not only depend on 
grain size but also the optical properties of the grain materials
prevents us from deriving more specific information on the actual grain size distribution 
from scattering and extinction observations alone.

\subsection{Non-Equilibrium Emission}
		It is likely that the lower size limit in the VSG population is set by 
interactions of VSGs with single photons in the interstellar radiation field. Given the 
density of this radiation field and the absorption cross sections of VSGs, a typical 
timescale between successive absorptions of energetic photons is about one day. It was 
recognized by Greenberg (1968), Duley
(1973), and Purcell (1976) that absorptions of energetic photons by systems
whose entire heat capacity is comparable to the energy of single photons will
lead to large temperature fluctuations. A typical maximum temperature of a particle 
consisting of N atoms, absorbing a 10eV photon while at zero temperature, can be estimated as
\begin{eqnarray}
 		T_{max}~\sim~ \frac{10~eV}{3Nk}~\sim~ \frac{39000}{N} [K]~\sim~ \frac{1}{a^{3}}.       
\end{eqnarray}
Thus, depending on the vaporization temperature of the material, VSGs with 
less than 40 to 50 atoms will not be stable against the absorption of single UV photons 
(Guhathakurta and Draine), and particles slightly larger will undergo temperature excursions
of 1000 K or more, resulting in temporary near- and mid-IR emission, generally referred to 
as non-equilibrium thermal emission (NETE). This is to be contrasted with the equilibrium 
thermal emission from the larger "typical" interstellar grains, which aquire a constant 
(low) temperature in interstellar space.
		First observational evidence for the presence of NETE in dusty astronomical 
sources was provided by Andriesse (1978) and Sellgren (1984), closely followed by the 
discovery by the Infrared Astronomy Satellite (IRAS) of the so-called infrared cirrus at 12 
and 25 microns (Boulanger et al. 1985), which was attributed to the same NETE process. 
Compared to earlier model predictions, which included only "typical" grains and SGs, thus 
terminating the size distribution at a = 5 nm, IRAS detected diffuse galactic background 
radiation five orders of magnitude more intense at 25 micron, and background
radiation twelve orders of magnitude more intense at 12 micron wavelength
(Puget \& Leger 1989).
 		From the analysis of data on the diffuse galactic infrared background 
radiation we know that VSGs absorb up to 40\% of the energy absorbed by dust in the Galaxy, 
but this ratio is highly dependent on the local environment and can vary by close to an 
order of magnitude for different clouds exposed to very similar radiation fields (Boulanger et al. 
1988). This suggests the relative abundance of VSG is highly variable in interstellar 
space. The fact that such variations appear totally uncorrelated with corresponding shape 
parameters for the UV extinction curves seen in the same directions suggests that the 
combined mass fraction of SGs and VSGs does not change but that the relative distribution 
between VSGs and SGs does. Efficient conglomeration of VSGs can remove them from the NETE 
regime without affecting the total mass in small grains and, consequently, the UV 
extinction.With the maximum temperature of VSGs being extremely sensitive to their size 
(Eqn. 3), the NETE process gives particularly clear information on VSGs near the lower 
limit of their size distribution and their relative abundance (Draine \& Anderson 1985; 
Weiland et al. 1986).

\subsection{Photoluminescence by Interstellar Grains}
		A second non-equilibrium emission process associated with interstellar 
grains is the so-called extended red emission (ERE). This process involves the absorption 
of UV/visible photons, followed by photoluminescence in a broad, featureless band, 
beginning at a wavelength of about 540 nm with a peak of maximum emission occurring at 
wavelengths ranging from 610 nm to 820 nm, depending on the environment where the ERE is 
produced. The ERE has been observed in a wide variety of sources, ranging from reflection 
nebulae, planetary nebulae, and HII regions, to dusty galaxies. Most important has been the 
recent detection of ERE in the diffuse interstellar medium of the Milky Way galaxy over a 
wide range of galactic latitudes by Gordon et al. (1998), for several reasons. First, it 
established the ERE phenomenon as a general characteristic of Galactic dust, not just a 
peculiar feature seen in special environments. Second, it allowed one to determine 
correlations between ERE intensities and atomic hydrogen column densities, which could then 
be converted to correlations to dust column densities. Thus, it was possible to determine
a lower limit to the ERE quantum efficiency of (10$\pm$3)\%, assuming that all
photons absorbed by interstellar grains in the 90 -- 550 nm wavelength range
are absorbed by the ERE-causing particles. This result implies that the true
ERE quantum efficiency is likely substantially larger than 10\%, because most absorption by 
interstellar grains is probably not caused by the luminescent variety. Also, the fact that 
10\% of all absorbed photons are contributing to the generation of the ERE implies that the 
ERE is produced by a fairly common grain component, derived from relatively abundant 
chemical elements.
		Examination of photoluminescence phenomena in nature in general
reveals that high efficiency luminescence occurs in only such systems where the exited 
electron is spatially confined and other modes of recombination are
effectively foreclosed. Examples are organic hydrocarbon molecules, dye molecules, or 
semiconductor nanoparticles. Witt et al. (1998) and Ledoux et al.
(1998) have advanced proposals that ERE is produced by a population of silicon nanoparticles
with about 3 nm diameter, while Seahra \& Duley (1999) have published a theoretical model 
attributing the ERE to photoluminescence by small carbon clusters. Earlier, d'Hendecourt et 
al. (1986) had suggested that polycyclic aromatic hydrocarbons might be the source of the 
ERE, but experimental data do not at present support such a proposal. In any event, if one 
adopts a likely intrinsic quantum efficiency of 50\%, as has been measured in
semiconductor nanoparticles and large molecular systems, the Gordon et al. (1998) result 
implies that about 20\% of the UV/optical photons absorbed in interstellar space are 
absorbed by luminescing VSGs.

\section{Summary}
The following main points have been reviewed:
\begin{enumerate}
\item Interstellar grains exist through a mass range of about $10^{-21}$ to
           $10^{-11}$ g. The lower limit of this range is determined by the  
	   stability of the particles against photodestruction resulting from
	   the absorption of single energetic photons. Only particles with 
	   mass in excess of $10^{-14}$ g are able to penetrate the heliopause.
\item Grains with radii larger than 100 nm contain most of the mass,
           grains smaller tham 100 nm contain most of the surface of 
	   interstellar particles.
\item UV extinction of star light provides evidence for the existence of
	   small grains (SGs and VSGs), but is insensitive to the size 
	   distribution of small grains.
\item The ratio of the mass of small grains (SGs and VSGs) to that of 
	   larger grains is highly variable in space, even within single cloud
	   complexes.
\item Data on scattering of star light confirm the absorption role of
	   small grains in the diffuse interstellar medium and dense nebulae.
\item Temperature fluctuations resulting from stochastic heating of very 
	   small grains and the resulting non-equilibrium thermal emission is
	   currently the strongest direct evidence for a large population of
	   nm-sized particles in the interstellar medium.
\item Large variations in relative abundances of these nanoparticles occur
	   without corresponding variations in the wavelength dependence of the
	   far-UV interstellar extinction, suggesting that nanoparticles mainly
	   conglomerate with other nanoparticles.
\item Photoluminescence (ERE) by grains in the diffuse interstellar medium
	   strongly suggests the presence of photo-luminescence-efficient 
	   nanoparticles or macro-molecules which absorb about 20\% of the 
	   UV/visible photons in the interstellar radiation field.
\end{enumerate}
\acknowledgments
		Constructive conversations with K.D. Gordon and D.G. Furton
are gratefully acknowledged. The author is particularly grateful to Priscilla
Frisch for invitations to the workshops on interstellar grains in the solar system at ISSI, 
Bern, and for her strong encouragement, without which this review would not have been 
written.

\end{document}